\documentclass[aps,prb,showpacs,twocolumn,superscriptaddress,10pt]{revtex4-1}
\usepackage{longtable}
\usepackage{amsfonts}
\usepackage{amsmath}
\usepackage{graphicx}
\usepackage{rotating}
\usepackage{dcolumn}
\newcolumntype{d}{D{.}{.}{3.5}}
\usepackage{subfloat}

\begin{document}
\author{Jacopo Baima}
\email{baima@insp.jussieu.fr}
\affiliation{ Sorbonne Universit\'e, CNRS, Institut des
  Nanosciences de Paris, UMR7588, F-75252, Paris, France}

\author{Francesco Mauri}
 \email{francesco.mauri@uniroma1.it}
 \affiliation{Dipartimento di Fisica, Universit\`{a} di Roma La Sapienza,
Piazzale Aldo Moro 5, I-00185 Roma, Italy and Graphene Labs, Fondazione Istituto Italiano di Tecnologia, Via Morego, I-16163 Genova, Italy}
  
\author{Matteo Calandra}
\email{calandra@insp.jussieu.fr}
\affiliation{ Sorbonne Universit\'e, CNRS, Institut des
  Nanosciences de Paris, UMR7588, F-75252, Paris, France}

\title{Field-effect-driven half-metallic multilayer graphene}
\date{\today}

\begin{abstract}
Rhombohedral stacked multilayer graphene displays the occurrence of a
 magnetic surface state at low temperatures. Recent angular resolved
 photoemission experiments demonstrate the robustness of the magnetic
 state in long sequences of ABC graphene. 
Here, by using first-principles calculations, 
we show that field-effect doping of these graphene multilayers 
induces a perfect half-metallic behaviour with 100\% of
 spin current polarization already at dopings attainable 
in conventional field effect transistors with solid state
dielectrics. 
Our work demonstrates the realisability of a new kind of  spintronic
devices where the transition between the low resistance and the high
resistance state is driven only by electric fields. 
\end{abstract}

\maketitle 
The realization of modern spintronic devices such as lateral spin-valves\cite{lateral_otani},
spin-hall effect devices and spin transfer torque 
memories\cite{SPtorque_Buhrman,SP_charge_Agnes} requires injection of  spin-polarized currents. 
Materials of choice for such devices 
are half-metals, namely compounds conducting in one spin channel and
insulating in the
other\cite{degoroot1983halfmetals,katsnelson2008halfmetals}, like Heussler alloys and 
some transition metal oxides\cite{katsnelson2008halfmetals}.
In many of these systems, however,
the surface conduction limits the amount of 
polarized current. 
Materials with 
a perfect half-metallic behavior not only in the bulk but also at the surface 
are needed. Half-metallicity embedded in surface or edge states of an easily 
processable material like carbon would be an ideal solution to the problem. 
However, carbon is not magnetic in its bulk forms.

Most of the search for half-metallic materials focused initially on 
ferromagnets and ferrimagnets, as in conventional antiferromagnets the 
cancellation of magnetic moments is due to symmetry operations connecting 
sites of opposite spin. The same symmetry relations cause the bands for
the two spin directions to be degenerate, forbidding a polarization of 
conduction electrons.\cite{hm-af-1} As a consequence, breaking the symmetry 
equivalence between sites is necessary to obtain a half-metallic material
 from an antiferromagnet. This can be done by substituting one of the magnetic 
centers in such a way that the net magnetic moment remains zero. The resulting
material is known as a spin compensated antiferromagnet and can 
be conceptualized as a special case of ferrimagnet with zero magnetic moment. 
Just as the ferrimagnetic case, the bands are not degenerate and 
half-metallicity can occur. \cite{hm-af-1,hm-af-2}

\begin{figure*}[t]\centering
 \includegraphics[width=1.999\columnwidth]{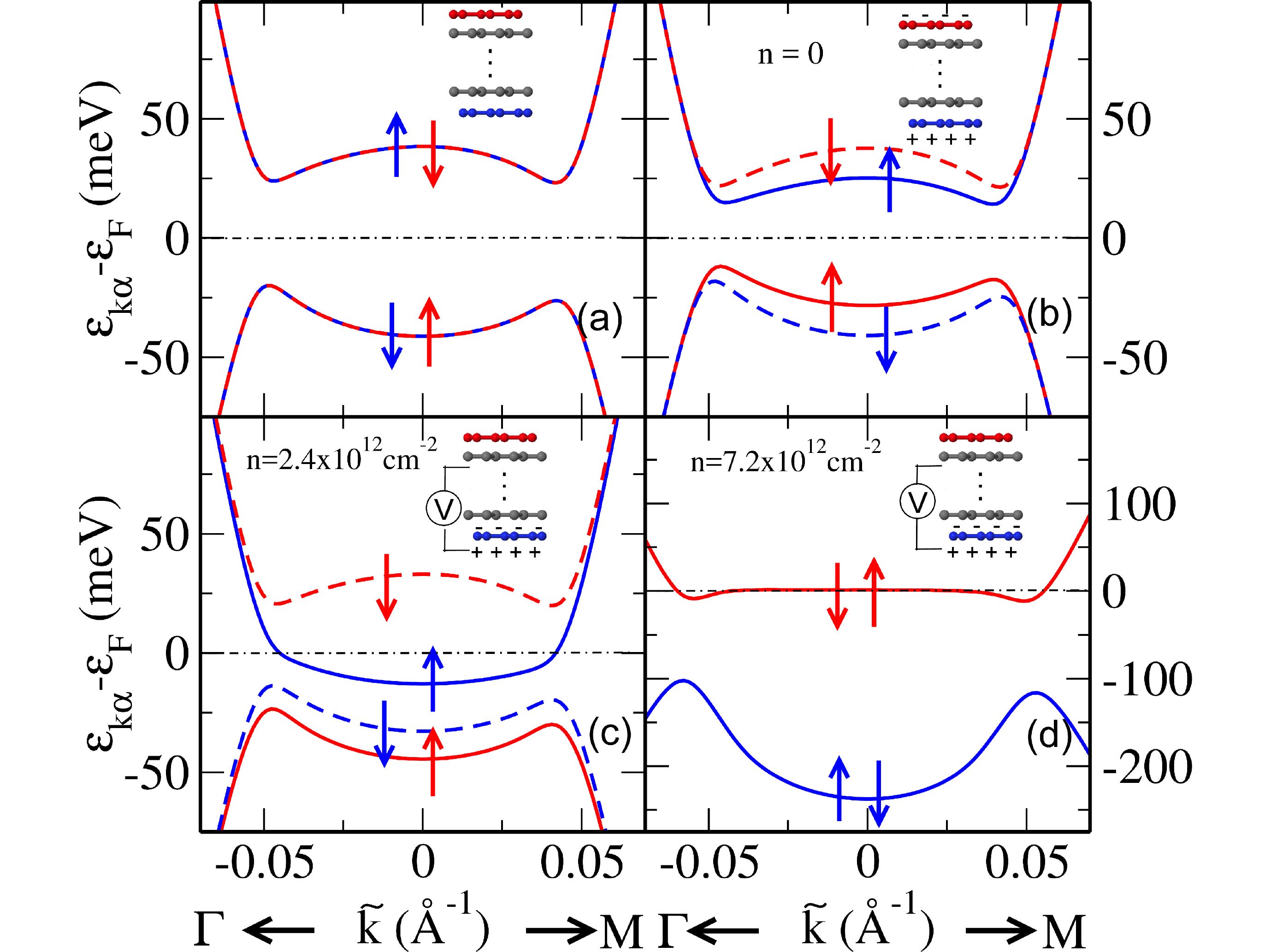}
 \caption{Calculated electronic structure using the PBE0\cite{abinitio} functional of
   rhombohedral 12-layer graphene close to  the K point
   ($\tilde{k}=k-\textup{K}$). In the insets we show a schematic view
   of the charged state considered in the calculation. The blue and
   red colors label the layers contributing to the surface bands.
The Fermi level is determined with a temperature $T=40$ K and labeled
   $\epsilon_F$. 
 (a)  Neutral antiferromagnetic ground
   state in the absence of external electric fields. (b) Electronic
   structure under the effect of an external
   electric field perpendicular to the layers in double gate
   configuration. 
  The electric field is
   simulated adding a layer of positive (negative) point charges at
   3.35\AA\ from the bottom (top) layer and
   keeping the multilayer graphene sample neutral, as shown in inset.
   The surface charge density for each gate is $\pm 0.8\times10^{12} \textup{cm}^{-2}$.
(c) and (d): electronic structure in field effect configuration at
 two different doping values $n$.
 In this case the system formed by the charged layer and the graphene
 multilayer is kept neutral (see insets). Continuous (dashed) lines label
 majority (minority) spin components. Blue (red) color label the dominant
 projection of the band structure over p$_z$ orbitals of the 
 atoms in the graphene layer close (opposite) to the charged layer.}
        \label{fig:undoped}
\end{figure*}

An alternative route is to introduce a symmetry-breaking perturbation 
in the antiferromagnetic material. This can lift the degeneracy while changing 
very little the net magnetization, 
with spins preserving an essentially antiferromagnetic configuration. 
Introducing such perturbation is possible if the spins of opposite
 orientation are spatially separated in the material. In the special case 
where they are localised at edges or surfaces, applying a static electric field 
is sufficient to break the symmetry and possibly induce a half-metallic state.

In a seminal paper, Son et al.\cite{son2006nanoribbons} showed 
that graphene nanoribbons have 
magnetic edge states and would turn into half-metallic ferromagnets 
under the application of an electric field. 
This finding has generated substantial interest,\cite{doi:10.1063/1.2908207,doi:10.1063/1.2821112,doi:10.1142/S1793292012500373,doi:10.1063/1.3143611,kim2008prediction,nanoribbon_fet}  and received some
preliminary experimental confirmation.\cite{joly2010nanorib} 
However, no such device 
has in practice been realized, since it would require application of  
very strong fields of the order of 0.1 V/\AA\  along the plane parallel to the 
nanoribbon\cite{son2006nanoribbons}.

 Previous theoretical work based on the Hubbard model applied to 
bilayer graphene\cite{bilayer_halfmetal}
suggested that an electric field combined wih doping could induce 
an half-metallic state. However, as the gap in bilayer graphene 
is only $\sim 2$ meV and the energy difference between up and down spin
states is of the same order, the magnetic state would be destroyed already at 10-20 K. For the same reason a low degree of spin
polarization can be expected in this system at finite temperatures.

A way to overcome these difficulties is to consider rhombohedral stacked multilayer graphene (RMG).
These systems present magnetic surface state and a sizeable gap, as large as 40 meV in RMG trilayers, more than
one order of magnitude larger than in the case of bilayer graphene.\cite{betul,lee2014,PhysRevB.86.115447,otherRMG,arpes} 
Rhombohedral or ABC stacking is 
a metastable phase of graphite which coexists with the 
more common AB (or Bernal) stacking.\cite{Lipson101}
Natural graphite samples contain
a substantial percentage of ABC graphite\cite{Lipson101,lui}. 
RMG can be obtained by exfoliation\cite{lee2014,henni} and identified through
Raman spectroscopy\cite{torche} or magneto-optic experiments\cite{henni}, as well as grown directly on a SiC substrate\cite{Pierucci}.

In this paper, by using first principles electronic structure
calculations\cite{abinitio}, we analyze the magnetic surface state of long series of RMG and show that
it becomes a half metal by doping the sample with the application
of a realistic electric field in a field-effect transistor (FET) configuration.
 We calculate amount of spin polarized conduction, finding that
doping charge densities as low as $2 \times 10^{12} {\rm cm}^{-2}$ are
sufficient to induce the half metallic behaviour
and to achieve a completely polarized spin current. 
At liquid nitrogen temperature the spin polarization is still 90\%. 

By neglecting spin polarization undoped RMG is a metal with an
extremely flat surface band at the Fermi level centered at the
special point K of the Brillouin zone\cite{betul}.  The surface state is formed by the $p_z$ orbitals of only one 
of the two atoms on each surface layer. This state is, however, unstable
in the spin channel, leading to a magnetic state and a gap opening at
low temperature, as shown in in Fig. \ref{fig:undoped} (a).
 Different magnetic instabilities are in principle possible,\cite{lee2014}
however the ground state was found to be globally
antiferromagnetic, with the atoms in the top graphene layer
having opposite spin polarizations with respect to those in the bottom one,
and ferrimagnetic within each layer.\cite{betul} 
 As a result the energy bands are twofold degenerate in the spin
 channel. The spin density decays 
rapidly inside the multilayer. 
The fraction of electrons contained in the surface state increases with increasing thickness,
as does the band gap which however saturates at 56 meV for 6
layers.\cite{betul} In thicker multilayers the gap decreases slowly,
approaching the band structure of bulk ABC graphite which shows Dirac points
along the $\Gamma$-K and K-M lines.

Transport measurements \cite{lee2014} on suspended ABC trilayers
confirm the opening of a 42 meV gap, in agreement with theoretical
calculations. The N{\'e}el critical temperature ($T_c$) as measured in
transport was found to be $34$ K.
More recently, ARPES experiments on thicker samples\cite{arpes}
($\sim$15 layers) demonstrated clearly the opening of a band gap, interpreted as due to 
magnetism, and a much higher $T_c$, larger than liquid nitrogen temperature.
Thus, both theoretical calculations and experimental data point to
a true antiferromagnet with antiferromagnetic in-plane and out-of-plane couplings.
For potential applications it is much more promising to consider
thick multilayers of RMG (6 to 20 layers) as the gap is larger, magnetism more robust
and the interaction with the substrate weaker. 

The inversion symmetry in RMG can be broken by applying an external electric field.  In
this case, the top and bottom graphene layers are not anymore
equivalent and the spin degeneracy is broken, as shown in
Fig. \ref{fig:undoped} (b). The system is still insulating, however,
the bottom of the conduction band is now formed by the up spin states
of the layer at the higher electrostatic potential and the top of the valence band
by the down spin states of the opposite layer.  Thus, applying moderate
electric fields breaks the spin degeneracy, reduces the gap but does
not the remove the insulating state and does not generate a
half-metallic state.  Much larger electric fields are necessary to
close the gap and form a metallic state in the spirit of
Ref. \onlinecite{son2006nanoribbons}.

 An alternative way to reach a half-metallic state is to doping RMG in a FET setup.
In this case, a gate dielectric is used and the RMG sample
is one of the two plates of the capacitor, as done in 
Refs. \onlinecite{schwierz2010graphene,radisavljevic2011single,Ye1193}. This is
shown
schematically in the inset of Figs. \ref{fig:undoped} (c) and (d).
The inversion symmetry is broken by the electric field
and a doping charge is introduced on the surface in contact with the dielectric.
This configuration can be simulated from first principles
by neglecting the gate dielectric and placing the system in front of a
charged layer, while doping it with the same amount of electrons obtaining
overall charge neutrality.\cite{fetsim}

At small doping, the extra charge is confined to the layer closest to
the gate (i.e. the layer of point charges in the insets of
Fig. \ref{fig:undoped} (c) and (d)) and decays inside
the multilayer very quickly.
The charge fills the unoccupied surface
band associated with the layer closest to the point charges. Because
unoccupied surface states of oppositely
oriented spins are localized at opposite surfaces, all the doping charge
will be spin polarized until the state located on the side of the gate 
is completely filled, generating a half-metallic state as shown in
Fig.   \ref{fig:undoped} (c).

\begin{figure}[!h]\centering 
  \includegraphics[width=0.99\columnwidth]{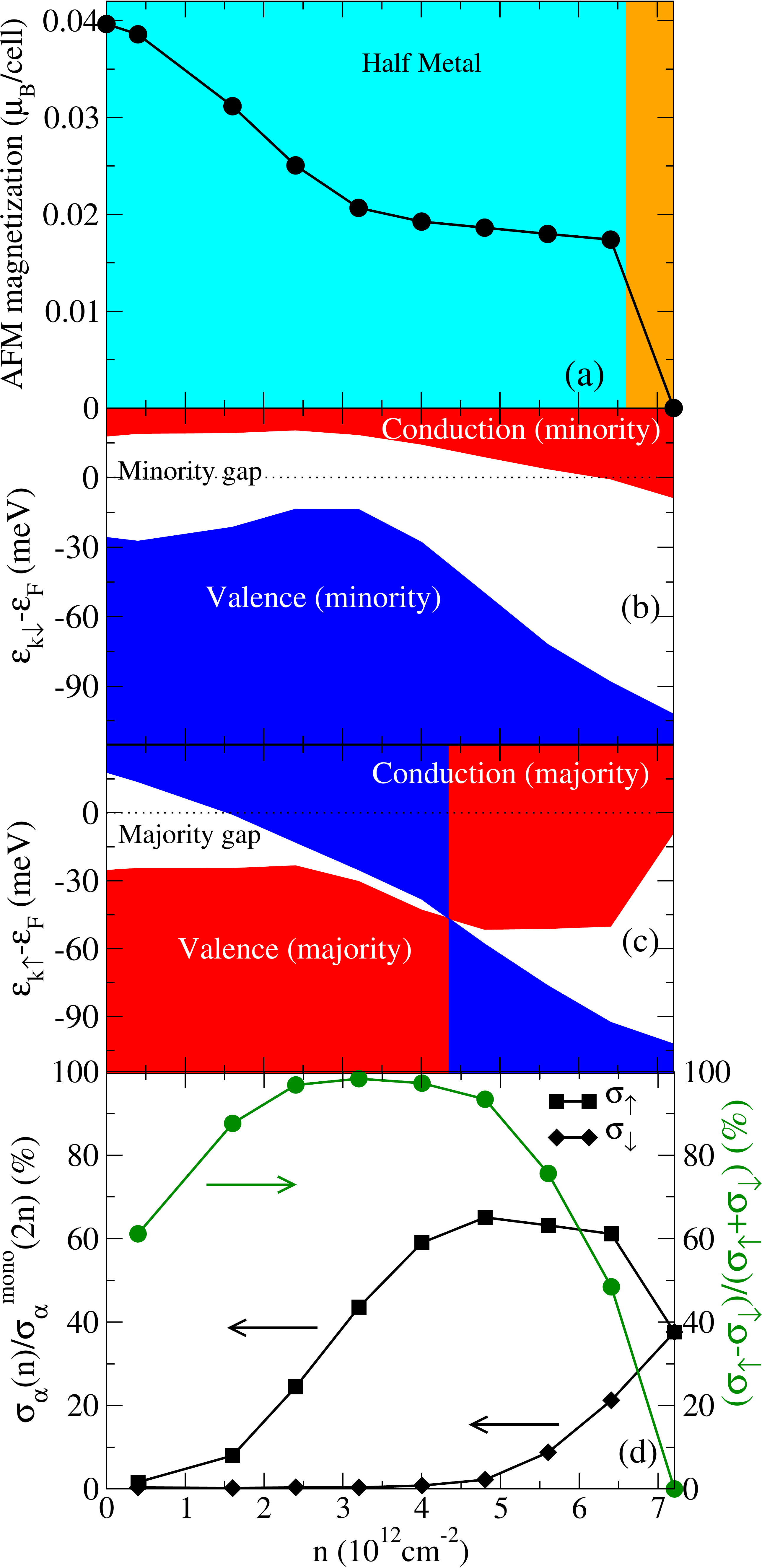}
                \caption{Magnetic and transport properties of 12-layer graphene 
                  with rhombohedral stacking as a function of 
                  field-effect doping at $T=40$ K. (a) Antiferromagnetic order 
                  parameter $\sum_i|m_i|$, where $m_i$ is the atomic 
                  magnetic moment. 
 (b) and (c) energy distance of the minority (majority) spin bands 
 from the Fermi level ($\epsilon_F$). 
The full colored regions  represents the dominant projection of
valence/conduction bands onto 
atomic states of the multilayer surfaces: blue for the layer closest to the 
charged layer, red for the opposite surface (see also Fig. \ref{fig:undoped}). 
(d) Spin polarized conduction with respect to the graphene 
conductivity at the same $k_F$ (left axis). The percentage of polarization 
of the conductivity is displayed on the right axis. 
}
        \label{fig:mainfig}
\end{figure}

For 12-layers, the
half-metallicity persists up to a doping charge density of about $6.4\times10^{12} \textup{cm}^{-2}$,
at which value the Fermi level crosses the spin down bands and
magnetism is lost, as shown in Fig. \ref{fig:mainfig} (a) and (b).
At larger doping the bands become again doubly
degenerate in the spin channel. 
The completely occupied electronic band is 
then formed by the $p_z$ states of one of the two atoms in 
the surface with higher electrostatic potential,
while the partially occupied one is formed by the  $p_z$ states
of one of the atoms in the opposite layer.
The half-metallic nature is lost and
the system becomes a metallic paramagnet, as shown in
Fig. \ref{fig:undoped} (d). 

The value of
critical doping at which half-metallicity disappears depends on the
number of electrons contained in the surface band, and as a consequence,
increases with multilayer thickness.\cite{betul}
Thicker samples result in flatter surface bands, that can in 
turn contain more electrons. 
Half metallicity is not only stronger in thicker samples but it 
also occurs in a larger range of doping and voltage. 

Half-metallicity is not, by itself, enough to obtain a completely
spin polarized current at finite temperature, as electrons could occupy
minority spin states above the Fermi level when thermally excited.
However, at a low doping level
the minority spin states closest to the Fermi level are at a distance of
$18$ meV $\sim 200$ K, as shown in Fig. \ref{fig:mainfig} (b). 
Nearly perfect spin filtering should then occur at low temperatures. 
Conversely, at liquid nitrogen or higher temperatures, a finite number of minority 
spin electrons (and a smaller number of holes) contributes to the conduction.
Within the constant relaxation-time approximation, $\tau({\bf k})=\tau(k_F)$
where $k_F$ is the Fermi momentum, the in-plane
conductivity for the spin channel $\alpha$ in the direction of a
transversal in-plane electric field
in the $x$ direction
can be expressed as:
\begin{align}\label{eq:conduct}
\sigma_{\alpha}=-\frac{\tau(k_F) e^2}{A N_k} \sum_{\mathbf{k},a}
  (v^{\mathbf{k} a \alpha}_x)^2 \left.\frac{\partial
  f}{\partial
  \epsilon}\right|_{\epsilon=\epsilon_{{\bf k}a\alpha} }
\end{align}
where $N_k$ is the number of k-points, $A$ is the surface of the 2D
unit cell, 
$f$ is the Fermi function, $\mathbf{v}^{\mathbf{k} a \alpha}=\frac{1}{\hbar}\partial \epsilon_{\mathbf{k} a \alpha}/\partial \mathbf{k}$ is the group velocity and $\tau(k_F)$ is the 
relaxation time, approximated here by a constant. 
This conductivity can be compared to that of monolayer graphene, namely
$\sigma_{\alpha}^{\rm mono}=\frac{e^2 \tau(k_F)}{2\pi
  \hbar^2} v_F k_F$ with $k_F=(\frac{n }{4\pi})^{\frac{1}{2}}$ 
and $\hbar v_F=7.75$ eV\AA\ , i. e. the PBE0 value, and $\beta=1/k_b T$.
Assuming that  n-doped monolayer graphene has similar scattering time to RMG
at the same $k_F$ (i.e. at twice larger doping in monolayer graphene than in
RMG), 
the ratio of $\sigma_{\alpha}(n)/\sigma_{\alpha}^{\rm mono}(2n)$ becomes
independent of $\tau$. 
This quantity is shown in the bottom panel of Fig. \ref{fig:mainfig}.
Despite the flattening of the bottom of the bands, 
in the doped case, the Fermi velocity is comparable to that of
monolayer graphene. This results in a sizeable conductivity in the
majority spin channel. On the contrary, in the minority spin channel
the velocity of the thermally activated electrons remain very small. 
The polarization of the conductivity $\frac{\sigma_\uparrow-\sigma_\downarrow}{\sigma_\uparrow+\sigma_\downarrow}$  
approaches 100\% around $3.2\times 10^{12} \text{cm}^{-2}$. 
For thinner multilayers, the unoccupied minority spin 
band is closer to the Fermi level, but the occupied one is further away. 
While the half-metallic state is less robust, holes do not contribute 
to the minority spin conductance. The results for 
6 layers RMG are provided in the Supplemental information\cite{supplemental}. The qualitative
behavior of the current polarization is similar to the 12 layer case but 
in a narrower range of doping, reaching a maximum of almost 100\% at 
$2.4\times 10^{12} \text{cm}^{-2}$. The maximum current polarization
  reduces with increasing temperature, however an high value of almost
  90\% could be obtained with liquid nitrogen cooling (see Supplemental information).\cite{supplemental}
 
Before concluding we discuss the
direction of the spin vectors, which is affected by both magnetic dipole and
spin-orbit interactions. Spin-orbit effects in carbon are small with
respect to the energy scales discussed in this paper, even considering terms
arising from interlayer interaction and broken symmetry due to the electric
field, \cite{graphene_spinorbit,bilayer_spinorbit} so they can be
safely neglected. 
The magnetic dipole-dipole interaction energy is
$E_{mag}\propto [\vec{\mu}_1\cdot\vec{\mu}_2- 3(\vec{\mu}_1\cdot\hat{r}_{12})(\vec{\mu}_2\cdot\hat{r}_{12})]/|r|^3$, and the main contribution 
comes from neighbouring spins within the layers, which are antiparallel.
In this case the second term favors
a spin configuration where the the magnetic moments are orthogonal to the vector
connecting them (the graphene bonds). As a consequence, the magnetic dipole
interaction will favour spin alignment perpendicular to the
layer itself. 

By using first principles electronic structure calculations, we have shown that 
field effect doping induces a perfect half-metallic behaviour  (100\%
spin polarization) in a range of  doping increasing with the film thicknesses. 
The main advantage of our proposal with respect to
previous attempts to develop carbon-based spintronic
devices\cite{son2006nanoribbons,bilayer_halfmetal} is that the half-metallic behaviour
occurs already at very low fields while surviving at least to liquid nitrogen temperature.
Conventional field effect transistor with solid state
dielectric can then be used to selectively populate one spin channel.
Finally, the similar character of  the valence and conduction bands
implies that 
a half metallic state occurs both
for p and n-doping.
However, changing the sign of the doping charge  implies a spin
inversion of the carriers.
As in FET it is possible to switch between p and
n-doping simply by changing the sign of the voltage across the
dielectric,  junctions between  field-effect p and n-doped RMG 
samples can be used to build spin valves in which the
transition between the low resistance and high resistance state is 
driven by electric fields and not by magnetic fields, as it normally
occurs.

We acknowledge support from the European Union Horizon 2020 research
and innovation 
program under Grant agreement No. 696656-GrapheneCore1.
Computer facilities were provided by CINES, IDRIS, and CEA TGCC (Grant
EDARI No. 2017091202).

%


\begin{thebibliography}{38}
\bibitem [{abi()}]{abinitio}%
  \BibitemOpen
  \href@noop {} {}\bibinfo {note} {We use the hybrid functional PBE0 within
  density functional theory (DFT) and an all-electron localized Gaussian basis
  set,\cite{crystal14} as this method was shown to provide reliable results for
  the stabilization of the magnetic state.\cite{betul} We use an electronic
  temperature of 40 K as it is well below the N{\'e}el point while being
  accessible to experiment. Sampling of reciprocal space is performed on mesh
  of 512x512 points, while the tolerance for the computation of direct-space
  electrostatic and exchange summations are set to 7 7 7 15 30.}\BibitemShut
  {Stop}%
\end{thebibliography}

\end{document}